\documentclass[nofootinbib, preprintnumbers, showpacs, twocolumn]{revtex4-1}
\usepackage{amsmath, amssymb, braket}
\usepackage[colorlinks = true]{hyperref}
\usepackage{graphicx}

\begin{document}
\preprint{KOBE-COSMO-17-03}
\title{Nonlinear resonant oscillation of gravitational potential\\
	induced by ultralight axion in $f(R)$ gravity}
\author{Arata Aoki}
\email{arata.aoki@stu.kobe-u.ac.jp}
\author{Jiro Soda}
\email{jiro@phys.sci.kobe-u.ac.jp}
\affiliation{Department of Physics, Kobe University, Kobe 657-8501, Japan}
\date{\today}
\pacs{
	95.35.+d,	
	98.62.Gq,	
	04.50.Kd	
}

\begin{abstract}
We study the ultralight axion dark matter with mass around $10^{-22}$\,eV in $f(R)$ gravity which might resolve the dark energy problem.
In particular, we focus on the fact that the pressure of the axion field oscillating in time produces oscillations of gravitational potentials.
We show that the oscillation of the gravitational potential is sensitive to the model of gravity.
Remarkably, we find that the detectability of the oscillation through the gravitational wave detectors can be significantly enhanced due to the nonlinear resonance between the ultralight axion and the scalaron.
\end{abstract}

\maketitle

\section{Introduction}
The cold dark matter (CDM) with a cosmological constant ($\Lambda$CDM model) is known as the standard cosmological model.
So far, supersymmetric particles, the so-called neutralinos, have been regarded as the most promising candidate for the CDM.
While the CDM works quite well on large scales, there exist problems on small scales.
In fact, the CDM model predicts an overabundance of structure on galactic scales, which is not consistent with observations.
Some common problems with CDM are the cusp-core problem~\cite{94:Flores, 94:Moore, 95:Burkert, 96:Navarro, 98:Moore, 04:Gentile}, the missing satellites problem~\cite{93:Kauffmann, 99:Klypin, 99:Moore}, and the too-big-to-fail problem~\cite{11:Boylan-Kolchin, 12:Boylan-Kolchin}.
Moreover, there was no signature of supersymmetry at the LHC.
Thus, it is worth investigating another possibility, namely the axion dark matter.

Originally, the axion was invented to resolve the strong CP problem in QCD~\cite{77:Peccei, 77:Peccei-2, 78:Weinberg, 78:Wilczek}.
Nowadays, however, it is known that the string theory also predicts axions with a wide range of mass scales~\cite{06:Svrcek, 10:Arvanitaki}.
Remarkably, the axion interacting very weakly with standard model particles is regarded as a candidate for the dark matter.
In particular, the ultralight axion with mass around $10^{-22}$\,eV can naturally resolve the problems on galactic scales because of its wave nature~\cite{00:Hu}.\footnote{
	Actually, there is a long history for the scalar dark matter.
	Some early works are Refs.~\cite{83:Baldeschi, 94:Sin, 96:Lee, 00:Sahni}.
	For complete references, see Ref.~\cite{16:Lee}.
}
Indeed, according to numerical simulations of dark matter halo density profiles with the ultralight scalar field, the mass around $10^{-22}$\,eV is favored by the data of dwarf spheroidal galaxies~\cite{12:Lora, 14:Schive, 16:Calabrese, 16:Gonzales-Morales}.
For this reason, the ultralight axion has recently attracted much attention.
Here, it would be fair to mention that the ``small scale problems of CDM" may be resolved even in the CDM framework by taking into account astrophysical processes including baryonic matter.

Remarkably, the pressure of the ultralight axion is oscillating in time with angular frequency at twice the axion mass, $\omega = 2m$.
Therefore, in order to find the axion dark matter, we should detect the oscillation of gravitational potential induced by this oscillating pressure.
Since the oscillation of the gravitational potential can be seen as a fluctuation of spacetime like gravitational waves, we would be able to detect the oscillation by means of gravitational-wave detectors.
Indeed, Khmelnitsky and Rubakov pointed out that the effect of oscillating pressure might be detected with pulsar timing array experiments~\cite{14:Khmelnitsky}.
We also pointed out that laser interferometer detectors can be used for this purpose~\cite{17:Aoki}.
One may think interferometers have no sensitivity to the isotropic oscillation of the pressure since its two arm lengths seem to change exactly the same amount.
However, the Solar System moves through the dark matter halo at a velocity about $v \sim 300 \, \text{km} / \text{s} = 10^{-3}$, and thus we feel the wind of the axion.
We see the axion wind as scalar gravitational waves, and the gravitational-wave interferometer detector does have sensitivity to the axion oscillation.

In addition to the dark matter, the dark energy is also a big issue in current physics.
In fact, the main energy component of the Universe is the dark energy.
Hence, it would be necessary to consider the detectability of the dark matter in the context of dark energy models.
Often, the cosmological constant is assumed when we discuss the detectability of dark matter.
However, the cosmological constant has several problems, e.g., the fine-tuning problem and coincidence problem.
One possibility to resolve these issues is to consider unknown matter such as the quintessence.
Unfortunately, there is no natural candidate for the quintessence in particle physics.
Therefore, it is worth investigating the possibility that the theory of gravity is different from Einstein gravity on cosmological scales.
Thus, we study the detectability of the ultralight axion dark matter in the context of modified gravity.

In our previous paper~\cite{16:Aoki}, we have discussed the detectability of the ultralight axion dark matter in the framework of $f(R)$ gravity, which is the simplest modified gravity.
We derived the gravitational potential sourced by the axion oscillation in the $f(R) \propto R^{2}$ model.
Remarkably, we found the resonance between the axion field and the scalaron field, which is the dynamical degree of freedom of $f(R)$ gravity, and the gravitational potential could be amplified dramatically.
However, it is not obvious if the resonance behavior can be seen in more general models.
In the $R^{2}$ model, the equation determining the gravitational potential is a linear equation under appropriate assumptions, while it should be nonlinear in more realistic models.
Hence, it is interesting to investigate whether general models can have resonance behavior.
In this paper, we will show that such a model does exist by constructing one specific model.

The paper is organized as follows:
In Sec.~II, we introduce a notion of axion dark matter and derive its energy-momentum tensor, which is a source term of the field equation for the metric.
In Sec.~III, we briefly summarize our previous work~\cite{16:Aoki}.
Then, we move on to the scalar-tensor formulation of $f(R)$ gravity and derive the formula (\ref{Eq_29}) for calculating the oscillating part of the gravitational potential.
In Sec.~IV, we discuss two specific models.
We show that the nonlinear resonance significantly enhances the amplitude of oscillation of the gravitational potential.
The final section is devoted to our conclusion.

\section{Axion Oscillation}
In this section, we introduce the axion dark matter.
In particular, we derive the energy-momentum tensor of the axion field, a source term of the metric field equation.

Let us assume the situation where the dark matter halo is composed of the ultralight axion.
Since the occupation number of the axion in the halo is huge, we can treat it as a classical scalar field.
The axion field satisfies the Klein-Gordon equation in the flat spacetime at the leading order, and the solution is given by the superposition of plane waves with different wave numbers and frequencies.
The wave number has a certain cutoff $k_{\text{max}} \sim mv$ due to the uncertainty principle, roughly set by the inverse of the de Broglie wavelength of an axion particle.
Since a typical velocity in the galaxy is $v \sim 10^{-3}$, we can assume that the axion field oscillates monochromatically with the angular frequency corresponding to its mass.
Under these assumptions, we can write the axion field as
\begin{equation}
	\phi(t, \vec{x}) = \phi_{0}(\vec{x})\cos[mt + \alpha(\vec{x})] \ ,
\end{equation}
where $\phi_{0}(\vec{x})$ is the amplitude and $\alpha(\vec{x})$ is the phase of the oscillation.
We can neglect the space dependence of $\phi_{0}(\vec{x})$ and $\alpha(\vec{x})$ at the leading order, and hereafter we will omit the phase $\alpha(\vec{x})$ for simplicity.

The energy density $\rho$ and the pressure $p$ of the axion field are given by
\begin{align}
	\rho &= \frac{1}{2}\dot{\phi}^{2} + \frac{1}{2}m^{2}\phi^{2} = \frac{1}{2}m^{2}\phi_{0}^{2} \equiv \rho_{0} \ , \\
	p &= \frac{1}{2}\dot{\phi}^{2} - \frac{1}{2}m^{2}\phi^{2} = -\rho_{0}\cos(2mt) \ .
\end{align}
The pressure oscillates in time with the angular frequency $\omega = 2m$.
Its amplitude is fixed by the local dark matter density $\rho_{0}$.
The energy-momentum tensor of the axion field is then given by
\begin{equation}
	T_{\mu\nu} = \begin{pmatrix} \rho_{0} & 0 \\ 0 & -\rho_{0}\cos(2mt) \delta_{{ij}} \end{pmatrix} \ .
\end{equation}
Finally, we obtain the trace of the energy-momentum tensor of the axion field as
\begin{equation}
	T = -\rho_{0}[1 + 3\cos(2mt)] \ .
	\label{1612042015}
\end{equation}
This is the source term of the field equation.
We can use Eq.~(\ref{1612042015}) as long as the axion field minimally couples to gravity.

We use the value $\rho_{0} = 0.3 \, \text{GeV} / \text{cm}^{3}$ as a typical energy density of the dark matter halo throughout the paper.\footnote{
	While this value is traditionally used, slightly higher values are reported in some papers, e.g., Refs.~\cite{10:Catena, 10:Salucci, 10:Pato, 11:McMillan, 12:Garbari}.
}
The period of the oscillation corresponds to about 1 year for $m = 10^{-22}$\,eV, and this time scale is much shorter than the cosmological time scale, i.e., $H_0^{-1} \sim 10^{10}$\,years.
Hence, after averaging the oscillating pressure over the cosmological time scale, the axion behaves as pressureless dust on cosmological scales.
Thus, the axion can be a candidate for the dark matter.

\section{Formula for Gravitational Potential Oscillation in $f(R)$ Gravity}
In this section, we will summarize our previous work~\cite{16:Aoki} and derive the formula (\ref{Eq_14}).
Then, we move on to the scalar-tensor formulation of $f(R)$ gravity and derive the main formula (\ref{Eq_29}) for calculating the time-dependent part of the gravitational potential in $f(R)$ gravity.

\subsection{Gravitational potential oscillation in $f(R)$ gravity}
The action for $f(R)$ gravity is given by
\begin{equation}
	S = \frac{1}{2} \int d^{4}x\sqrt{-g} \, [R + f(R)] + S_{\text{m}} \ ,
	\label{Eq_6}
\end{equation}
where $f(R)$ is a function of the Ricci scalar $R$ and $S_{\text{m}}$ is the action for matter fields.
We assume $f(R) \ll R$ and $f_{R} \equiv f'(R) \ll 1$ so that the deviation from Einstein's theory is small.
Taking the variation of the action with respect to the metric, we obtain the metric field equation:
\begin{equation}
	G_{\mu\nu} - \frac{1}{2}g_{\mu\nu}f + (R_{\mu\nu} + g_{\mu\nu}\Box - \nabla_{\mu}\nabla_{\nu})f_{R} = T_{\mu\nu} \ ,
\end{equation}
where $G_{\mu\nu} \equiv R_{\mu\nu} - (1 / 2)g_{\mu\nu}R$ is the Einstein tensor, $T_{\mu\nu}$ is the energy-momentum tensor for the matter field, and $\Box \equiv \nabla_{\mu}\nabla^{\mu}$ is the d'Alembert operator.
The trace of the field equation reads
\begin{equation}
	3\Box f_{R} - R + Rf_{R} - 2f = T \ .
\end{equation}
We assume that the spatial derivative of $f_{R}$ is much smaller than the time derivative of it, i.e., $\Box f_{R} \simeq -\ddot{f}_{R}$.
This is because the length scale of the dark matter halo, about 10\,kpc or larger, is much larger than the time scale of the oscillation, $m^{-1} \sim 0.1$\,pc for $m = 10^{-22}$\,eV.
We use this kind of approximation throughout the paper.
Under the assumption, the field equation becomes
\begin{equation}
	3\ddot{f}_{R} + R = -T \ ,
	\label{Eq_9}
\end{equation}
where we used $f \ll R$ and $f_{R}\ll 1$.
Although we assumed $f_{R} \ll 1$, the contribution of $f_{R}$ to the equation of motion is not so small.
This is because the $f_{R}$ term appears as $\ddot{f}_{R} \sim (2m)^{2}f_{R}$ in Eq.~(\ref{Eq_9}).
Thus, the $f_{R}$ term is enhanced by a factor of $m^{2} / R_{0} = m^{2} / \rho_{0} \sim 10^{17}(m / 10^{-22} \, \text{eV})^{2}$ and can be comparable to the Ricci scalar $R$ in Eq.~(\ref{Eq_9}).
Hereafter, we consider the axion as the matter field.
Since the axion field minimally couples to gravity in $f(R)$ gravity, we can use Eq.~(\ref{1612042015}) for the energy-momentum tensor of the axion field.

Since the gravitational potentials are small even in the dark matter halo, they can be treated as perturbations.
Let us use the Newtonian gauge for the metric:
\begin{equation}
	g_{\mu\nu} = \begin{pmatrix} -1 - 2\Psi & 0 \\ 0 & (1 - 2\Phi)\delta_{ij} \end{pmatrix} \ .
\end{equation}
Note that the expansion of the Universe is completely negligible on the scale of the dark matter halo.
At the first order of the potentials, the Ricci scalar is calculated as
\begin{equation}
	R = -6\ddot{\Phi} + 2\nabla^{2}(2\Phi - \Psi) \ .
\end{equation}

Let us write the Ricci scalar as the sum of the time-independent part $R_{0}$ and the time-dependent part $\delta R$:
\begin{equation}
	R = R_{0} + \delta R \ ,
\end{equation}
where $R_{0}$ is defined as the long-term average of $R$, $R_{0} \equiv \braket{R}$.
We also separate the gravitational potential $\Phi$ ($\Psi$) into the time-independent part $\Phi_{0} \equiv \braket{\Phi}$ ($\Psi_{0} \equiv \braket{\Psi}$) and the time-dependent part $\delta\Phi$ ($\delta\Psi$).
We have the equation $\Phi_{0} = \Psi_{0}$ from the traceless part of the space-space component of the Einstein equation.
Hence, $R_{0}$ can be written as $R_{0} = 2\nabla^{2}\Phi_{0}$.
The field equation (\ref{Eq_9}) gives $R_{0} \equiv \braket{R} = \rho_{0}$, and this is nothing but the Poisson equation
\begin{equation}
	2\nabla^{2}\Phi_{0} = \rho_{0} \ .
\end{equation}
Assuming $|\delta\ddot{\Phi}| \gg |\nabla^{2}\delta\Phi|$ and $|\delta\ddot{\Phi}| \gg |\nabla^{2}\delta\Psi|$, $\delta R$ is approximately given by
\begin{equation}
	\delta R = -6\delta\ddot{\Phi} \ .
\end{equation}
Integrating this twice and using the field equation (\ref{Eq_9}), we obtain
\begin{equation}
	\delta\Phi = \frac{\rho_{0}}{8m^{2}}\cos(2mt) + \frac{1}{2}(f_{R} - \braket{f_{R}}) \ ,
	\label{Eq_14}
\end{equation}
where $\braket{f_{R}}$ is the average value of $f_{R}$.
This is the formula for calculating the time-dependent part of the gravitational potential.
Therefore, in order to obtain the amplitude of the gravitational potential, we first solve the field equation (\ref{Eq_9}) and then substitute the solution into Eq.~(\ref{Eq_14}).

For Einstein theory, we have $f_{R} = 0$.
Thus, Eq.~(\ref{Eq_14}) gives
\begin{equation}
	\delta\Phi = \delta\Phi_{\text{E}}\cos(2mt) \ ,
	\label{Eq_15}
\end{equation}
where
\begin{equation}
	\delta\Phi_{\text{E}} \equiv \frac{\rho_{0}}{8m^{2}} = 5 \times 10^{-18} \left( \frac{10^{-22} \, \text{eV}}{m} \right)^{2} \ .
\end{equation}
The frequency of the gravitational potential is
\begin{equation}
	f = \frac{2m}{2\pi} = 5 \times 10^{-8} \, \text{Hz} \, \left( \frac{m}{10^{-22} \, \text{eV}} \right) \ .
\end{equation}
This is consistent with the result derived by Khmelnitsky and Rubakov~\cite{14:Khmelnitsky}.

\subsection{Formula for gravitational potential oscillation}
While we have discussed the axion oscillation in $f(R)$ gravity, it is convenient to move on to the equivalent scalar-tensor theory for qualitative understanding of the physics~\cite{10:Sotiriou, 10:DeFelice, 11:Nojiri}.
In this section, we will reformulate $f(R)$ gravity in terms of the scalar-tensor theory.

The action for $f(R)$ gravity can be rewritten into that of the scalar-tensor theory as
\begin{equation}
	S = \frac{1}{2} \int d^{4}x\sqrt{-g} \, \big[ (1 + \varphi)R - 2U(\varphi) \big] + S_{\text{m}} \ ,
	\label{Eq_18}
\end{equation}
where $U(\varphi)$ is defined by
\begin{equation}
	U(\varphi) \equiv \frac{1}{2} \left[ \varphi A(\varphi) - f \big( A(\varphi) \big) \right] \ ,
	\label{Eq_19}
\end{equation}
and $A(\varphi)$ is the solution of
\begin{equation}
	f'(A) = \varphi \ .
	\label{Eq_20}
\end{equation}
Taking the variation of the action (\ref{Eq_18}) with respect to $\varphi$, we obtain the constraint
\begin{equation}
	R = 2U_{,\varphi} = A(\varphi) \ .
\end{equation}
Substituting this into the action (\ref{Eq_18}), we obtain the original $f(R)$ action (\ref{Eq_6}).
As long as $f''(A) \not= 0$, the solution of Eq.~(\ref{Eq_20}) is uniquely determined and thus the two theories are completely equivalent.
The variation of the action (\ref{Eq_18}) with respect to the metric gives the field equation:
\begin{equation}
	(1 + \varphi)G_{\mu\nu} + g_{\mu\nu}U(\varphi) + \left( g_{\mu\nu}\Box - \nabla_{\mu}\nabla_{\nu} \right) \varphi = T_{\mu\nu} \ .
\end{equation}
The trace of this equation gives the equation of motion for the scalar field $\varphi$:
\begin{equation}
	3\Box\varphi - A(\varphi) + \varphi A(\varphi) - 2f \big( A(\varphi) \big) = T \ ,
\end{equation}
where we used Eq.~(\ref{Eq_19}).
Hence, we can interpret $\varphi$, which is often called a scalaron, as an extra dynamical degree of freedom in the theory.
Let us define the potential of the scalar field $V(\varphi)$ by
\begin{equation}
	V_{,\varphi} \equiv \frac{1}{3} \big[ A(\varphi) - \varphi A(\varphi) + 2f \big( A(\varphi) \big) \big] \ .
\end{equation}
Using the approximations $f(A) \ll A$ and $\varphi = f'(A) \ll 1$, the derivative of the potential $V_{,\varphi}$ is approximated by
\begin{equation}
	V_{,\varphi} \simeq \frac{A(\varphi)}{3} \ .
\end{equation}
Hereafter we will use this approximate form for $V_{,\varphi}$.
Under the assumption $\Box\varphi \simeq -\ddot{\varphi}$, the equation of motion is written as
\begin{equation}
	\ddot{\varphi} + V_{,\varphi} = \frac{\rho_{0}}{3} \big[ 1 + 3\cos(2mt) \big] \ ,
	\label{Eq_26}
\end{equation}
where we substituted the energy-momentum tensor (\ref{1612042015}).
Let us introduce the effective potential $V_{\text{eff}}(\varphi)$ by
\begin{equation}
	V_{\text{eff},\varphi} \equiv V_{,\varphi} - \frac{\rho_{0}}{3} = \frac{1}{3}(A(\varphi) - \rho_{0}) \ .
	\label{Eq_27}
\end{equation}
The effective potential has a minimum at $R = A(\varphi) = \rho_{0}$ as desired.
Using the effective potential, the equation of motion is written in a simple form:
\begin{equation}
	\ddot{\varphi} + V_{\text{eff},\varphi} = \rho_{0}\cos(2mt) \ .
	\label{Eq_28}
\end{equation}
Since $\varphi = f_{R}$, Eq.~(\ref{Eq_14}) can be rewritten in terms of the scalar field $\varphi$ as
\begin{equation}
	\delta\Phi = \delta\Phi_{\text{E}}\cos(2mt) + \frac{1}{2}(\varphi - \braket{\varphi}) \ ,
	\label{Eq_29}
\end{equation}
where $\delta\Phi_{\text{E}} \equiv \rho_{0} / 8m^{2}$.
Therefore, in order to obtain the time-dependent part of the gravitational potential in the scalar-tensor formulation, we first calculate the effective potential from a function $f(R)$ by using Eq.~(\ref{Eq_27}), solve the equation of motion (\ref{Eq_28}), and then substitute the solution into the formula (\ref{Eq_29}).

\section{Nonlinear Resonant Oscillation}
In this section, we will study two specific models.
First, we illustrate the resonance of oscillation using the $R^2$ model.
Then, we show that the exponential model exhibits nonlinear resonance.

\subsection{$R^{2}$ model}
We first consider the $f(R) \propto R^{2}$ model, which can be solved analytically.
We have already studied this model in the previous paper in the framework of $f(R)$ gravity~\cite{16:Aoki}.
However, since the model is useful for qualitative understanding of general $f(R)$ models, here we again discuss the same model in the scalar-tensor theory.

The model is defined by
\begin{equation}
	f(R) = \frac{R^{2}}{6M^{2}} \ ,
\end{equation}
where $M$ is the mass scale of the model, which is independent of the Ricci scalar in this model.
The effective potential is calculated as
\begin{equation}
	V_{\text{eff}}(\varphi) = \frac{1}{2}M^{2}(\varphi - \varphi_{0})^{2} \ ,
\end{equation}
where $\varphi_{0} \equiv f'(\rho_{0}) = \rho_{0} / 3M^{2}$.
Hence, the mass scale $M$ is nothing but the mass of the scalaron field $\varphi$.
We expect this model to capture some features of general models since any analytic function can be approximated by a quadratic function near its minimum.
The equation of motion for $\varphi$ now is a linear equation:
\begin{equation}
	\ddot{\varphi} + M^{2}(\varphi - \varphi_{0}) = \rho_{0}\cos(2mt) \ .
\end{equation}
The solution is given by
\begin{equation}
	\varphi = \varphi_{0} + \frac{\rho_{0}}{M^{2} - (2m)^{2}}\cos(2mt) \ .
\end{equation}
Here we focus on the induced solution by the axion oscillation and omitted the homogeneous solutions.
This is in part because the homogeneous solutions decay in the expanding Universe by the Hubble friction, and we expect that only the induced solution remains in the present Universe~\cite{07:Starobinsky}.
Since the long-time average of $\varphi$ is $\braket{\varphi} = \varphi_{0}$, Eq.~(\ref{Eq_29}) gives
\begin{equation}
	\delta\Phi = \frac{\delta\Phi_{\text{E}}}{1 - (2m / M)^{2}}\cos(2mt) \ .
	\label{Eq_34}
\end{equation}
Let us introduce a parameter $\mu$ as
\begin{equation}
	\mu \equiv \frac{2m}{M} \ .
\end{equation}
When $\mu \ll 1$, the amplitude of $\delta\Phi$ becomes $\delta\Phi_{\text{amp}} \simeq \delta\Phi_{\text{E}}$, and the result (\ref{Eq_15}) is recovered.
In the opposite case $\mu \gg 1$, the amplitude is suppressed as $\delta\Phi_{\text{amp}} \simeq (1 / \mu^{2})\delta\Phi_{\text{E}}$.
Remarkably, when $\mu \simeq 1$, the resonance occurs and the gravitational potential is amplified.
The behavior near the resonance point should be strongly dependent on the details of models.
To see this, we study another model in the next subsection.

\subsection{Exponential model}
In the $R^{2}$ model, the equation of motion for $\varphi$ is linear.
In general $f(R)$ models, however, the equation of motion for $\varphi$ should be nonlinear.
In the nonlinear cases, it is not obvious whether the resonance occurs.
Indeed, in the previous paper~\cite{16:Aoki}, we saw that there are no stable resonance solutions near $\mu = 1$ in the Hu-Sawicki model~\cite{07:Hu} and the Starobinsky model~\cite{07:Starobinsky}.
In this subsection, we will show that there exist nonlinear models with resonance behavior by constructing one specific model.
We will also see that the resonance phenomena acquires new interesting properties when nonlinearity is taken into account.

Let us consider the following model:
\begin{equation}
	f(R) = \frac{R_{0}^{2}}{3\lambda^{2}M^{2}}\exp \left[ -\lambda \left( \frac{R}{R_{0}} - 1 \right) \right] \ ,
	\label{Eq_36}
\end{equation}
where $M$ is the mass scale of the model at $R = R_{0}$, i.e., $M^{2} = 1 / 3f''(R_{0})$, and $\lambda$ is a positive parameter.
The exponential-type model was first introduced by Ref.~\cite{08:Cognola}.
The effective potential is calculated as
\begin{equation}
	V_{\text{eff}}(\varphi) = \frac{R_{0}}{3\lambda}\varphi \left[ 1 - \ln \left( \frac{\varphi}{\varphi_{0}} \right) \right] \ ,
\end{equation}
where $\varphi_{0}$ is the field value at the minimum of the effective potential:
\begin{equation}
	\varphi_{0} \equiv f'(R_{0}) = -\frac{R_{0}}{3\lambda M^{2}} \ .
\end{equation}
The model satisfies the solar system constraint if the following condition is held~\cite{10:DeFelice}:
\begin{equation}
	|\varphi_{0}| = \frac{3 \times 10^{-18}}{\lambda} \left( \frac{2 \times 10^{-22} \, \text{eV}}{M} \right)^{2} < 3 \times 10^{-15} \ .
\end{equation}
Hence, the model can pass the solar system test even when $M = 2 \times 10^{-22} \, \text{eV}$ if $\lambda > 10^{-3}$.
Of course, the larger the mass scale $M$ is, the easier the model passes the solar system test.

\begin{figure}
\includegraphics[width=.6\hsize]{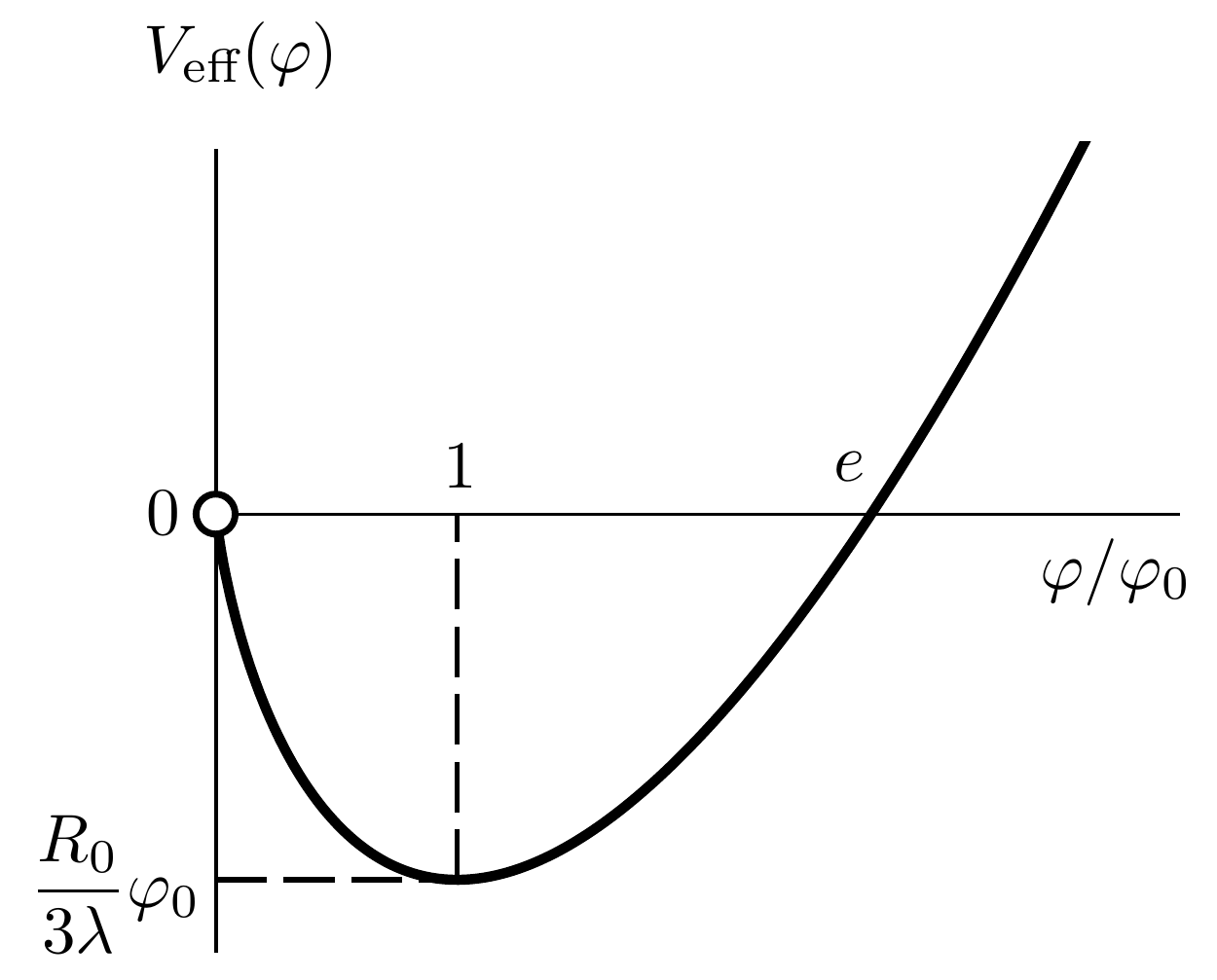}
\caption{
	The functional form of the effective potential $V_{\text{eff}}(\varphi)$ in the exponential model (\ref{Eq_36}).
	$V_{\text{eff}}(\varphi)$ has a minimum at $\varphi = \varphi_{0}$.
	The point $\varphi = 0$ is a singularity, at which the gradient of the potential diverges.
	Since $\varphi = 0$ corresponds to $R \to +\infty$, this is called the curvature singularity.
}
\label{Fig_1}
\end{figure}

We plot the functional form of the effective potential in Fig.~\ref{Fig_1}.
The effective potential has a singularity at $\varphi = 0$, where $R = A(\varphi) = 3V_{,\varphi}$ diverges.
This kind of singularity is known as a curvature singularity~\cite{07:Briscese, 08:Nojiri, 08:Frolov, 08:Kobayashi}, which often appears in $f(R)$ models.
We can always remove the singularity without affecting the dynamics on the halo scale by adding a regularization term, e.g., $R^{2} / 6\mathcal{M}^{2}$ with a large $\mathcal{M}$~\cite{08:Nojiri, 08:Dev, 09:Kobayashi, 10:Appleby}.
However, we have no need of such a modification for our purpose to construct a concrete model with the resonance behavior.
Therefore, we will not discuss the regularization and focus on what the model (\ref{Eq_36}) predicts.
In this model the scalar field can practically move only within the range
\begin{equation}
	0 < \varphi / \varphi_{0} \lesssim e \ .
\end{equation}
This is because if we start with $V_{\text{eff}}(\varphi) > 0$, i.e., $\varphi / \varphi_{0} > e$, the scalar field easily hits the curvature singularity.
Hence possible amplitudes of the scalar field are roughly limited to $(e / 2)|\varphi_{0}| \sim |\varphi_{0}|$, and the maximum amplitude of the gravitational potential is approximated as
\begin{equation}
	\delta\Phi_{\text{max}} \sim \frac{1}{2}|\varphi_{0}| = \frac{R_{0}}{6\lambda M^{2}} \sim \frac{1}{\lambda} \times 10^{-17} \left( \frac{10^{-22} \, {\text{eV}}}{M} \right)^{2} \ ,
	\label{Eq_40}
\end{equation}
where we assumed the second term in Eq.~(\ref{Eq_29}) is dominant.
The amplitude could become large for sufficiently small $\lambda$ even when we fix $\mu \sim 1$, i.e., $M \sim 2m$.

Now, we will show that the solutions with the maximum amplitude (\ref{Eq_40}) do exist.
All we have to do is to solve the nonlinear equation of motion
\begin{equation}
	\bar{\varphi}'' + \ln\bar{\varphi} = - 3\lambda\cos(\mu\tau) \ ,
\end{equation}
where we used the dimensionless quantities $\tau = Mt$, $\mu = 2m / M$, $\bar{\varphi} = \varphi / \varphi_{0}$, and the prime denotes a derivative with respect to $\tau$.
In the case of a linear forced oscillator, the general solution is given by the superposition of an induced solution and homogeneous solutions.
However, in nonlinear cases, the superposition of two solutions no longer gives a solution.
Therefore, it is hard to solve nonlinear systems analytically.
Indeed, we have to rely on a perturbative method or numerical calculations.

\begin{figure*}
\includegraphics[width=.8\hsize]{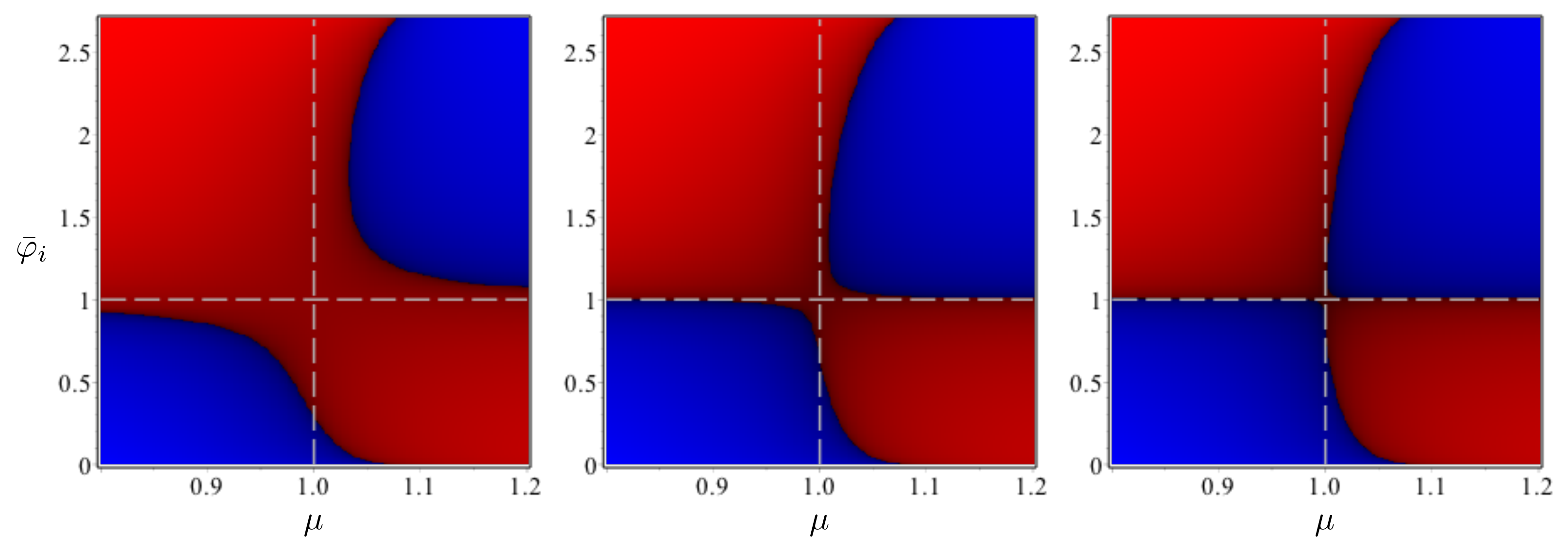}
\caption{
	Plot of $\bar{\varphi}'(T / 2)$ in the $(\mu, \bar{\varphi}_{i})$ plane for $\lambda = 10^{-2}, 10^{-3}, 10^{-4}$ from left to right.
	The red (blue) regions correspond to positive (negative) value of $\bar{\varphi}'(T / 2)$.
	Since $\bar{\varphi}'(T / 2)$ is a smooth function of $(\mu, \bar{\varphi}_{i})$, there exist the boundaries with $\bar{\varphi}'(T / 2) = 0$, which correspond to closed-orbit solutions.
}
\label{Fig_2}
\end{figure*}

First, in order to get an intuition, we use the following perturbative method.
The deviation from the minimum $\chi$ is defined by $\bar{\varphi} =1 + \chi$.
When $\chi \ll 1$, we can approximate the equation by
\begin{equation}
	\chi'' + \chi - \frac{1}{2}\chi^{2} + \frac{1}{3}\chi^{3} = - 3\lambda\cos(\mu\tau) \ .
	\label{Eq_42}
\end{equation}
This can be analyzed perturbatively.
First, we seek a resonant solution around $\mu \sim 1$.
To this aim, we rewrite Eq.~(\ref{Eq_42}) as
\begin{equation}
	\chi'' + \mu^{2}\chi = (\mu^{2} -1)\chi + \frac{1}{2}\chi^{2} - \frac{1}{3}\chi^{3} - 3\lambda\cos(\mu\tau) \ .
	\label{Eq_43}
\end{equation} 
We regard the terms in the right-hand side as small perturbations.
Then, we can solve Eq.~(\ref{Eq_43}) with the expansion
\begin{equation}
	\chi = \chi_{0} + \chi_{1} + \cdots \ .
\end{equation}
Substituting the series into Eq.~(\ref{Eq_43}), we obtain the lowest order solution
\begin{equation}
	\chi_{0} = A\cos(\mu\tau) \ .
\end{equation}
At the next order, we have secular sources
\begin{align}
	\chi''_{1} + \mu^{2}\chi_{1} &= (\mu^{2} - 1)\chi_{0} + \frac{1}{2}\chi_{0}^{2} - \frac{1}{3}\chi_{0}^{3} - 3\lambda\cos(\mu\tau) \nonumber\\
		&= \left\{ (\mu^{2} - 1)A - 3\lambda - \frac{1}{4}A^{3} \right\} \cos(\mu\tau) + \cdots \ .
\end{align}
If this secular term remains, we will have a secular solution with the growing amplitude.
In other words, we have to renormalize this secular evolution into the frequency.
This can be achieved by imposing the condition
\begin{equation}
	(A^{2} - 4\mu^{2} + 4)A = - 12\lambda \ .
\end{equation}
Apparently, this allows the order one solution $A \simeq 2\sqrt{\mu^{2} -1}$ for $\mu \geq 1$.
Thus, we have shown that there exists the resonant oscillation with the amplitude close to the maximum one.
In principle, we can repeat the same analysis to find other resonant solutions.
We will soon show explicit examples with the numerical method.

Next, we investigate the same system numerically.
We seek periodic solutions satisfying $(\bar{\varphi}(T), \bar{\varphi}'(T)) = (\bar{\varphi}(0), \bar{\varphi}'(0))$, where $T \equiv 2\pi / \mu$ is the period of the oscillation.
In other words, we look for the solutions with closed orbit in the phase space $(\bar{\varphi}, \bar{\varphi}')$.
Note that in the $R^{2}$ model, this condition removes homogeneous solutions.
In order to find such solutions, in general, we should study the map $(\bar{\varphi}(0), \bar{\varphi}'(0)) \mapsto (\bar{\varphi}(T), \bar{\varphi}'(T))$ and find its fixed points.
However, since the equation of motion (\ref{Eq_26}) now has the time reflection symmetry $t \to -t$ thanks to the absence of a friction term in addition to the time translation symmetry $t \to t + T$, the orbit is closed if $\bar{\varphi}'(T / 2) = 0$ when starting with the initial condition $(\bar{\varphi}(0), \bar{\varphi}'(0)) = (\bar{\varphi}_{i}, 0)$.
Note that the reverse statement is not always true; there exist periodic solutions that do not satisfy $\bar{\varphi}'(T / 2) = 0$ as will be seen later.

\begin{figure}
\includegraphics[width=.7\hsize]{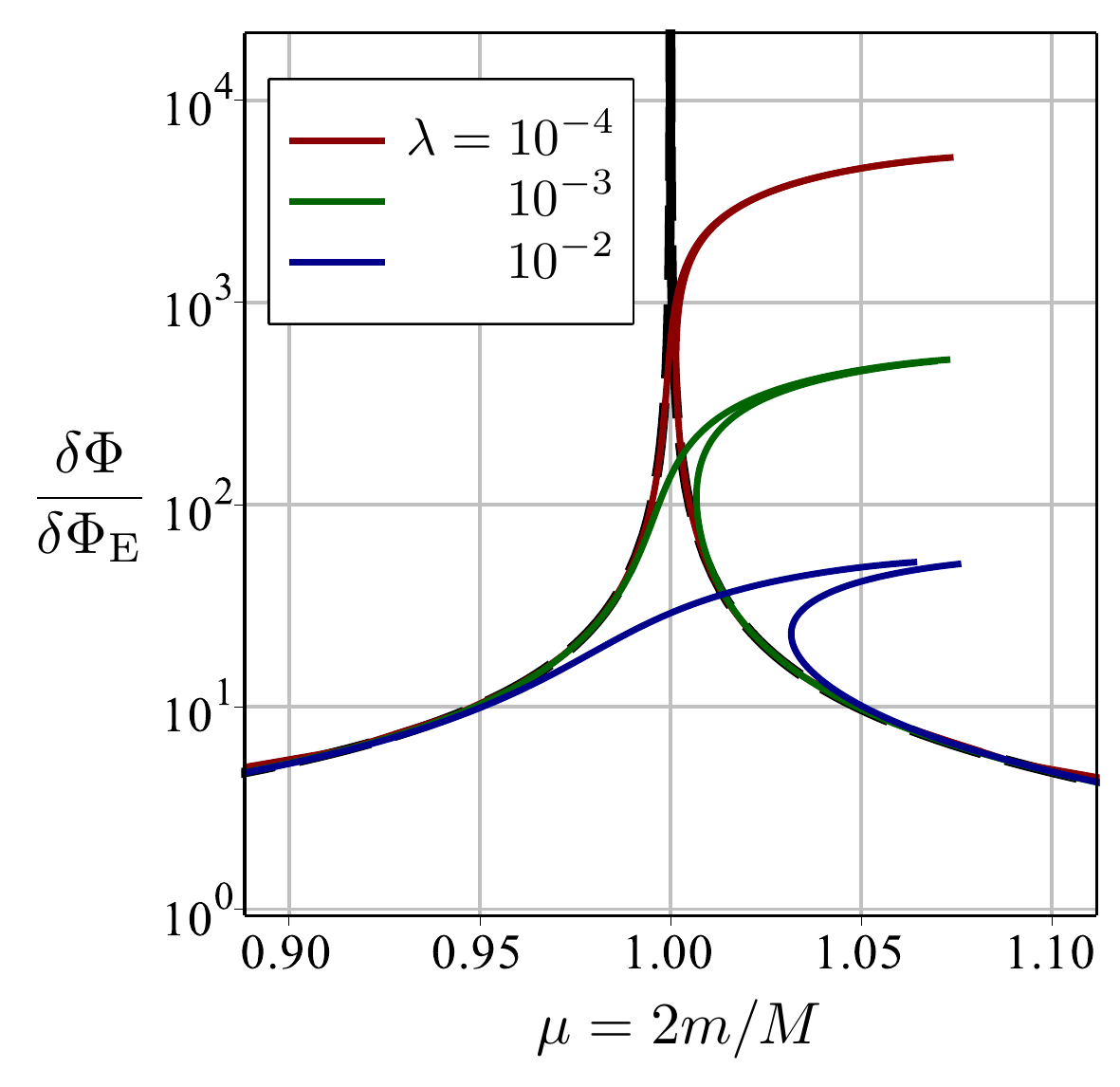}
\caption{
	The resonance curves near the resonance point with different values of $\lambda$.
	The black dashed line is the amplitude in the $R^{2}$ model, Eq.~(\ref{Eq_34}).
	The resonance curves in the exponential model cannot be distinguished from those in the $R^{2}$ model except for $\mu \sim 1$.
	The maximum amplitude is inversely proportional to $\lambda$ as seen in Eq.~(\ref{Eq_40}).
	The discontinuity of the curve seen in the case $\lambda = 10^{-2}$ comes from the absence of a friction term, as in the case of the $R^{2}$ model.
}
\label{Fig_3}
\end{figure}

In Fig.~\ref{Fig_2}, we plot the value $\bar{\varphi}'(T / 2)$ as a function of $(\mu, \bar{\varphi}_{i})$ with different values of $\lambda$.
The red (blue) regions correspond to positive (negative) values of $\bar{\varphi}'(T / 2)$.
Since $\bar{\varphi}'(T / 2)$ is a smooth function of $(\mu, \bar{\varphi}_{i})$, there are boundaries with $\bar{\varphi}'(T / 2) = 0$ between two regions, which correspond to closed-orbit solutions.
Hence, Fig.~\ref{Fig_2} shows that solutions exist with amplitude $|\bar{\varphi}| = |\varphi / \varphi_{0}| \sim \mathcal{O}(1)$.
The figure also shows that there are three solutions around $\mu \gtrsim 1$.
This is a general feature of nonlinear forced oscillators~\cite{60:Landau}.
Which solution is selected should be determined by initial conditions.
We also show the corresponding resonance curves in Fig.~\ref{Fig_3}.
The resonance curves are bent by nonlinearity compared to the $R^{2}$ case.
This plot shows that the resonance curves in the exponential model cannot be distinguished from those in the $R^{2}$ model except for $\mu \sim 1$.
Namely, while the behavior near the resonance point is strongly dependent on models, the $R^{2}$ model extracts some features of general models as desired.

\begin{figure}
\includegraphics[width=.6\hsize]{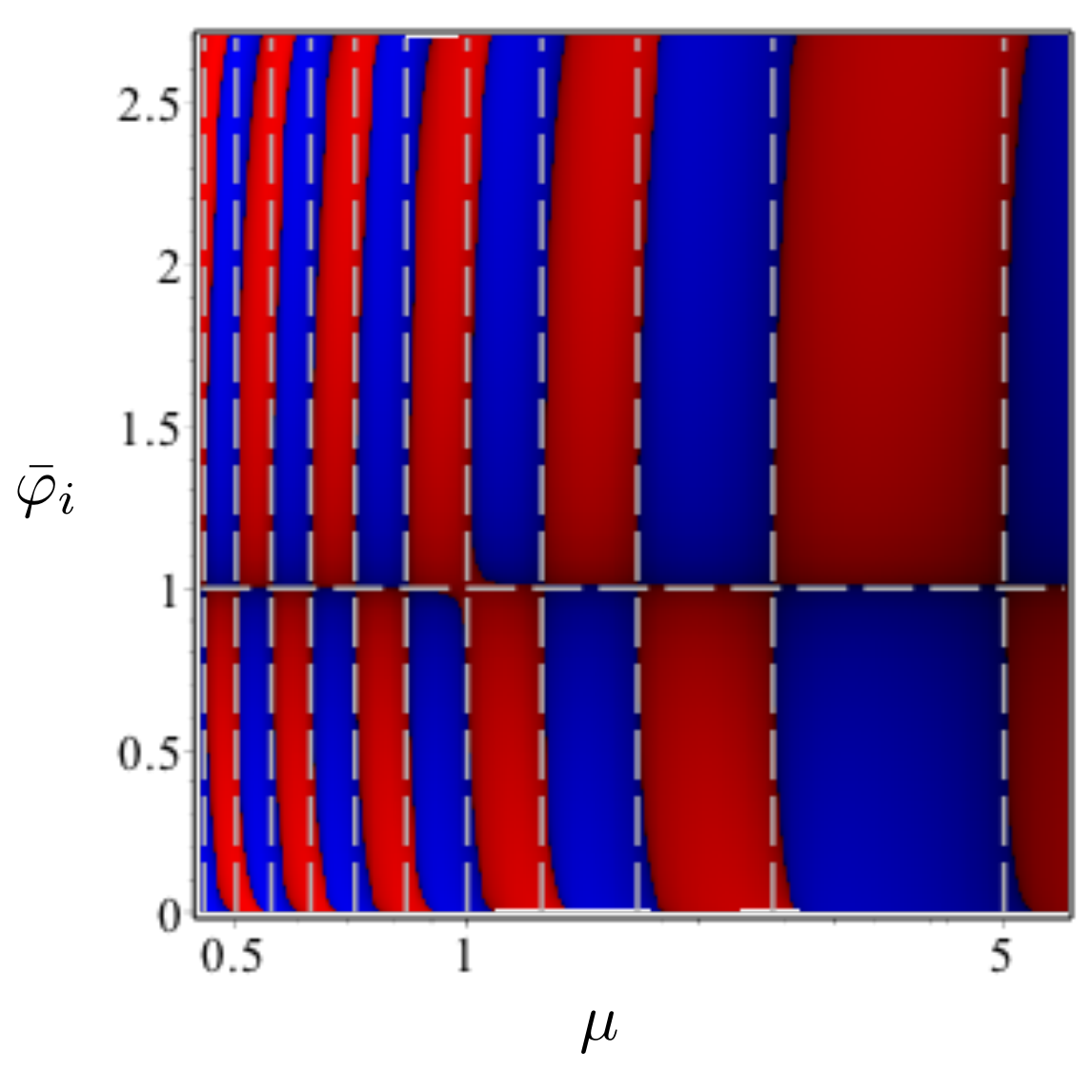}
\caption{
	Resonance points with $\mu = q / 5~(q = 1, 2, 3 \dots)$ for $\lambda = 10^{-3}$.
	The plot shows the existence of multiple resonance points in the case of the nonlinear model.
}
\label{Fig_4}
\end{figure}

Remarkably, in the nonlinear case, there appear new resonances at $\mu = q / p$ for positive integers $p$ and $q$~\cite{60:Landau}.
As an example, we show the same plot as Fig.~\ref{Fig_2} for the series with $p = 5$ in Fig.~\ref{Fig_4}, where we chose $\lambda = 10^{-3}$.
Note that the plotted function $\bar{\varphi}'(T / 2)$ is replaced by $\bar{\varphi}'(5T / 2)$ in Fig.~\ref{Fig_4}.
In these cases, the resonance of scalaron field oscillation occurs when the relation $ M\sim 2m(p /q)$ holds.
Hence, the observed frequency becomes higher or lower than $2m$.
In the former cases, the frequencies could be accessible by ground-based gravitational-wave detectors such as LIGO.

\section{Conclusion}
In this paper, we have studied the axion oscillation in $f(R)$ gravity.
The point was that the pressure of the axion field oscillating in time produces the oscillation of the gravitational potential.
We have derived the formula (\ref{Eq_29}) for calculating the time-dependent part of the gravitational potential.
It turns out that the amplitude of the oscillation of the gravitational potential is determined by the amplitude of the scalar field (scalaron) in the scalar-tensor formulation.
Remarkably, the amplitude could be amplified dramatically by the resonance when the mass of the scalaron is sufficiently close to twice the axion mass.
We have also shown that there appear subharmonics resonances at $\mu = q / p$ in the case of nonlinear models by studying one specific model~(\ref{Eq_36}) as an example.

In this paper, we have studied two $f(R)$ models on halo scales.
We should mention that we did not attempt to construct complete models which also explain the accelerated expansion of the present Universe.
In fact, in order to explain the dark energy, we need to modify the models on cosmological scales.
Since the cosmological critical density is much lower than the dark matter density in the halo, $\rho_{\text{cr}} / \rho_{0} \sim 10^{-5}$, we expect that the modification of the functional form of $f(R)$ on cosmological scales does not affect the dynamics on halo scales.

The oscillating gravitational potential can, in principle, be observed by using gravitational-wave detectors.
So far, two ideas for direct detection of the ultralight axion dark matter through the gravitational interaction have been proposed: pulsar timing array experiments~\cite{14:Khmelnitsky} and laser interferometers~\cite{17:Aoki}.\footnote{
	Recently, a new detection method with binary pulsars was also proposed~\cite{16:Blas}.
	The nonlinear resonance that we found in this paper further enhanced the detectability of the ultralight axion.
}

Although we have focussed on the present Universe, there may be other phenomena caused by the oscillating axion field.
In particular, we should revisit the structure formation process in the presence of the ultralight axion field.
Structure formation in modified gravity is often discussed assuming the validity of static or quasistatic approximation on galactic scales.
However, if the ultralight axion is the dark matter, the situation is quite different.
We should take into account the oscillating pressure of the axion field when we discuss structure formation in modified gravity.

\acknowledgments
This work was in part supported by JSPS Grant-in-Aid for JSPS Research Fellow Grant No. 17J00568 (A.A.), JSPS KAKENHI Grant No. 17H02894, and MEXT KAKENHI Grant No. 15H05895 (J.S.).

\bibliographystyle{myBst}
\bibliography{draft}

\begin{thebibliography}{10}

\bibitem{94:Flores}R.~A. Flores and J.~R. Primack, ``{\em {Observational and
  theoretical constraints on singular dark matter
  halos}},"\,\href{http://dx.doi.org/10.1086/187350}{Astrophys. J. {\bfseries
  427}, L1 (1994)},
  \href{http://arxiv.org/abs/astro-ph/9402004}{arXiv:astro-ph/9402004}.
\bibitem{94:Moore}B.~Moore, ``{\em {Evidence against dissipation-less dark
  matter from observations of galaxy
  haloes}},"\,\href{http://dx.doi.org/10.1038/370629a0}{Nature {\bfseries 370},
  629 (1994)}.
\bibitem{95:Burkert}A.~Burkert, ``{\em {The Structure of Dark Matter Halos in
  Dwarf Galaxies}},"\,\href{http://dx.doi.org/10.1086/309560}{Astrophys. J.
  {\bfseries 447}, L25 (1995)},
  \href{http://arxiv.org/abs/astro-ph/9504041}{arXiv:astro-ph/9504041}.
\bibitem{96:Navarro}J.~F. Navarro, C.~S. Frenk, and S.~D.~M. White, ``{\em {The
  Structure of Cold Dark Matter
  Halos}},"\,\href{http://dx.doi.org/10.1086/177173}{Astrophys. J. {\bfseries
  462}, 563 (1996)},
  \href{http://arxiv.org/abs/astro-ph/9508025}{arXiv:astro-ph/9508025}.
\bibitem{98:Moore}B.~Moore, F.~Governato, T.~R. Quinn, J.~Stadel, and G.~Lake,
  ``{\em {Resolving the Structure of Cold Dark Matter
  Halos}},"\,\href{http://dx.doi.org/10.1086/311333}{Astrophys. J. {\bfseries
  499}, L5 (1998)},
  \href{http://arxiv.org/abs/astro-ph/9709051}{arXiv:astro-ph/9709051}.
\bibitem{04:Gentile}G.~Gentile, P.~Salucci, U.~Klein, D.~Vergani, and
  P.~Kalberla, ``{\em {The cored distribution of dark matter in spiral
  galaxies}},"\,\href{http://dx.doi.org/10.1111/j.1365-2966.2004.07836.x}{Mon.
  Not. Roy. Astron. Soc. {\bfseries 351}, 903 (2004)},
  \href{http://arxiv.org/abs/astro-ph/0403154}{arXiv:astro-ph/0403154}.
\bibitem{93:Kauffmann}G.~Kauffmann, S.~D.~M. White, and B.~Guiderdoni, ``{\em
  {The Formation and Evolution of Galaxies Within Merging Dark Matter
  Haloes}},"\,\href{http://dx.doi.org/10.1093/mnras/264.1.201}{Mon. Not. Roy.
  Astron. Soc. {\bfseries 264}, 201 (1993)}.
\bibitem{99:Klypin}A.~A. Klypin, A.~V. Kravtsov, O.~Valenzuela, and F.~Prada,
  ``{\em {Where Are the Missing Galactic
  Satellites?}},"\,\href{http://dx.doi.org/10.1086/307643}{Astrophys. J.
  {\bfseries 522}, 82 (1999)},
  \href{http://arxiv.org/abs/astro-ph/9901240}{arXiv:astro-ph/9901240}.
\bibitem{99:Moore}B.~Moore, S.~Ghigna, F.~Governato, G.~Lake, T.~R. Quinn,
  J.~Stadel, and P.~Tozzi, ``{\em {Dark Matter Substructure within Galactic
  Halos}},"\,\href{http://dx.doi.org/10.1086/312287}{Astrophys. J. {\bfseries
  524}, L19 (1999)},
  \href{http://arxiv.org/abs/astro-ph/9907411}{arXiv:astro-ph/9907411}.
\bibitem{11:Boylan-Kolchin}M.~Boylan-Kolchin, J.~S. Bullock, and M.~Kaplinghat,
  ``{\em {Too big to fail? The puzzling darkness of massive Milky Way
  subhaloes}},"\,\href{http://dx.doi.org/10.1111/j.1745-3933.2011.01074.x}{Mon.
  Not. Roy. Astron. Soc. {\bfseries 415}, L40 (2011)},
  \href{http://arxiv.org/abs/1103.0007}{arXiv:1103.0007 [astro-ph.CO]}.
\bibitem{12:Boylan-Kolchin}M.~Boylan-Kolchin, J.~S. Bullock, and M.~Kaplinghat,
  ``{\em {The Milky Way's bright satellites as an apparent failure of
  $\Lambda$CDM}},"\,\href{http://dx.doi.org/10.1111/j.1365-2966.2012.20695.x}{Mon.
  Not. Roy. Astron. Soc. {\bfseries 422}, 1203 (2012)},
  \href{http://arxiv.org/abs/1111.2048}{arXiv:1111.2048 [astro-ph.CO]}.
\bibitem{77:Peccei}R.~D. Peccei and H.~R. Quinn, ``{\em {CP Conservation in the
  Presence of
  Pseudoparticles}},"\,\href{http://dx.doi.org/10.1103/PhysRevLett.38.1440}{Phys.
  Rev. Lett. {\bfseries 38}, 1440 (1977)}.
\bibitem{77:Peccei-2}R.~D. Peccei and H.~R. Quinn, ``{\em {Constraints imposed
  by $\mathrm{CP}$ conservation in the presence of
  pseudoparticles}},"\,\href{http://dx.doi.org/10.1103/PhysRevD.16.1791}{Phys.
  Rev. D {\bfseries 16}, 1791 (1977)}.
\bibitem{78:Weinberg}S.~Weinberg, ``{\em {A New Light
  Boson?}},"\,\href{http://dx.doi.org/10.1103/PhysRevLett.40.223}{Phys. Rev.
  Lett. {\bfseries 40}, 223 (1978)}.
\bibitem{78:Wilczek}F.~Wilczek, ``{\em {Problem of Strong $P$ and $T$
  Invariance in the Presence of
  Instantons}},"\,\href{http://dx.doi.org/10.1103/PhysRevLett.40.279}{Phys.
  Rev. Lett. {\bfseries 40}, 279 (1978)}.
\bibitem{06:Svrcek}P.~Svrcek and E.~Witten, ``{\em {Axions in string
  theory}},"\,\href{http://dx.doi.org/10.1088/1126-6708/2006/06/051}{J. High
  Energy Phys. {\bfseries 06}, 051 (2006)},
  \href{http://arxiv.org/abs/hep-th/0605206}{arXiv:hep-th/0605206}.
\bibitem{10:Arvanitaki}A.~Arvanitaki, S.~Dimopoulos, S.~Dubovsky, N.~Kaloper,
  and J.~March-Russell, ``{\em {String
  axiverse}},"\,\href{http://dx.doi.org/10.1103/PhysRevD.81.123530}{Phys. Rev.
  D {\bfseries 81}, 123530 (2010)},
  \href{http://arxiv.org/abs/0905.4720}{arXiv:0905.4720 [hep-th]}.
\bibitem{00:Hu}W.~Hu, R.~Barkana, and A.~Gruzinov, ``{\em {Fuzzy Cold Dark
  Matter: The Wave Properties of Ultralight
  Particles}},"\,\href{http://dx.doi.org/10.1103/PhysRevLett.85.1158}{Phys.
  Rev. Lett. {\bfseries 85}, 1158 (2000)},
  \href{http://arxiv.org/abs/astro-ph/0003365}{arXiv:astro-ph/0003365}.
\bibitem{83:Baldeschi}M.~R. Baldeschi, R.~Ruffini, and G.~B. Gelmini, ``{\em
  {On massive fermions and bosons in galactic
  halos}},"\,\href{http://dx.doi.org/10.1016/0370-2693(83)90688-3}{Phys. Lett.
  B {\bfseries 122}, 221 (1983)}.
\bibitem{94:Sin}S.-J. Sin, ``{\em {Late-time phase transition and the galactic
  halo as a Bose
  liquid}},"\,\href{http://dx.doi.org/10.1103/PhysRevD.50.3650}{Phys. Rev. D
  {\bfseries 50}, 3650 (1994)},
  \href{http://arxiv.org/abs/hep-ph/9205208}{arXiv:hep-ph/9205208}.
\bibitem{96:Lee}J.-w. Lee and I.-g. Koh, ``{\em {Galactic halos as boson
  stars}},"\,\href{http://dx.doi.org/10.1103/PhysRevD.53.2236}{Phys. Rev. D
  {\bfseries 53}, 2236 (1996)},
  \href{http://arxiv.org/abs/hep-ph/9507385}{arXiv:hep-ph/9507385}.
\bibitem{00:Sahni}V.~Sahni and L.~Wang, ``{\em {New cosmological model of
  quintessence and dark
  matter}},"\,\href{http://dx.doi.org/10.1103/PhysRevD.62.103517}{Phys. Rev. D
  {\bfseries 62}, 103517 (2000)},
  \href{http://arxiv.org/abs/astro-ph/9910097}{arXiv:astro-ph/9910097}.
\bibitem{16:Lee}J.-W. Lee, ``{\em {Characteristic size and mass of galaxies in
  the Bose-Einstein condensate dark matter
  model}},"\,\href{http://dx.doi.org/10.1016/j.physletb.2016.03.016}{Phys.
  Lett. B {\bfseries 756}, 166 (2016)},
  \href{http://arxiv.org/abs/1511.06611}{arXiv:1511.06611 [astro-ph.GA]}.
\bibitem{12:Lora}V.~Lora, J.~Magana, A.~Bernal, F.~J. S\'{a}nchez-Salcedo, and
  E.~K. Grebel, ``{\em {On the mass of ultra-light bosonic dark matter from
  galactic
  dynamics}},"\,\href{http://dx.doi.org/10.1088/1475-7516/2012/02/011}{J.
  Cosmol. Astropart. Phys. {\bfseries 1202}, 011 (2012)},
  \href{http://arxiv.org/abs/1110.2684}{arXiv:1110.2684 [astro-ph.GA]}.
\bibitem{14:Schive}H.-Y. Schive, T.~Chiueh, and T.~Broadhurst, ``{\em {Cosmic
  structure as the quantum interference of a coherent dark
  wave}},"\,\href{http://dx.doi.org/10.1038/nphys2996}{Nat. Phys. {\bfseries
  10}, 496 (2014)}, \href{http://arxiv.org/abs/1406.6586}{arXiv:1406.6586
  [astro-ph.GA]}.
\bibitem{16:Calabrese}E.~Calabrese and D.~N. Spergel, ``{\em {Ultra-light dark
  matter in ultra-faint dwarf
  galaxies}},"\,\href{http://dx.doi.org/10.1093/mnras/stw1256}{Mon. Not. Roy.
  Astron. Soc. {\bfseries 460}, 4397 (2016)},
  \href{http://arxiv.org/abs/1603.07321}{arXiv:1603.07321 [astro-ph.CO]}.
\bibitem{16:Gonzales-Morales}A.~X. Gonz\'{a}les-Morales, D.~J.~E. Marsh,
  J.~Pe\~{n}arrubia, and L.~Ure\~{n}a L\'{o}pez, ``{\em {Unbiased constraints
  on ultralight axion mass from dwarf spheroidal
  galaxies}},"\,\href{http://arxiv.org/abs/1609.05856}{arXiv:1609.05856
  [astro-ph.CO]}.
\bibitem{14:Khmelnitsky}A.~Khmelnitsky and V.~Rubakov, ``{\em {Pulsar timing
  signal from ultralight scalar dark
  matter}},"\,\href{http://dx.doi.org/10.1088/1475-7516/2014/02/019}{J. Cosmol.
  Astropart. Phys. {\bfseries 1402}, 019 (2014)},
  \href{http://arxiv.org/abs/1309.5888}{arXiv:1309.5888 [astro-ph.CO]}.
\bibitem{17:Aoki}A.~Aoki and J.~Soda, ``{\em {Detecting ultralight axion dark
  matter wind with laser
  interferometers}},"\,\href{http://dx.doi.org/10.1142/S0218271817500638}{Int.
  J. Mod. Phys. D {\bfseries 26}, 1750063 (2017)},
  \href{http://arxiv.org/abs/1608.05933}{arXiv:1608.05933 [astro-ph.CO]}.
\bibitem{16:Aoki}A.~Aoki and J.~Soda, ``{\em {Pulsar timing signal from
  ultralight axion in $f(R)$
  theory}},"\,\href{http://dx.doi.org/10.1103/PhysRevD.93.083503}{Phys. Rev. D
  {\bfseries 93}, 083503 (2016)},
  \href{http://arxiv.org/abs/1601.03904}{arXiv:1601.03904 [hep-ph]}.
\bibitem{10:Catena}R.~Catena and P.~Ullio, ``{\em {A novel determination of the
  local dark matter
  density}},"\,\href{http://dx.doi.org/10.1088/1475-7516/2010/08/004}{J.
  Cosmol. Astropart. Phys. {\bfseries 08}, 004 (2010)},
  \href{http://arxiv.org/abs/0907.0018}{arXiv:0907.0018 [astro-ph.CO]}.
\bibitem{10:Salucci}P.~Salucci, F.~Nesti, G.~Gentile, and C.~F. Martins, ``{\em
  {The dark matter density at the Sun's
  location}},"\,\href{http://dx.doi.org/10.1051/0004-6361/201014385}{Astron.
  Astrophys. {\bfseries 523}, A83 (2010)},
  \href{http://arxiv.org/abs/1003.3101}{arXiv:1003.3101 [astro-ph.GA]}.
\bibitem{10:Pato}M.~Pato, O.~Agertz, G.~Bertone, B.~Moore, and R.~Teyssier,
  ``{\em {Systematic uncertainties in the determination of the local dark
  matter density}},"\,\href{http://dx.doi.org/10.1103/PhysRevD.82.023531}{Phys.
  Rev. D {\bfseries 82}, 023531 (2010)},
  \href{http://arxiv.org/abs/1006.1322}{arXiv:1006.1322 [astro-ph.HE]}.
\bibitem{11:McMillan}P.~J. McMillan, ``{\em {Mass models of the Milky
  Way}},"\,\href{http://dx.doi.org/10.1111/j.1365-2966.2011.18564.x}{Mon. Not.
  Roy. Astron. Soc. {\bfseries 414}, 2446 (2011)},
  \href{http://arxiv.org/abs/1102.4340}{arXiv:1102.4340 [astro-ph.GA]}.
\bibitem{12:Garbari}S.~Garbari, C.~Liu, J.~I. Read, and G.~Lake, ``{\em {A new
  determination of the local dark matter density from the kinematics of K
  dwarfs}},"\,\href{http://dx.doi.org/10.1111/j.1365-2966.2012.21608.x}{Mon.
  Not. Roy. Astron. Soc. {\bfseries 425}, 1445 (2012)},
  \href{http://arxiv.org/abs/1206.0015}{arXiv:1206.0015 [astro-ph.GA]}.
\bibitem{10:Sotiriou}T.~P. Sotiriou and V.~Faraoni, ``{\em {$f(R)$ theories of
  gravity}},"\,\href{http://dx.doi.org/10.1103/RevModPhys.82.451}{Rev. Mod.
  Phys. {\bfseries 82}, 451 (2010)},
  \href{http://arxiv.org/abs/0805.1726}{arXiv:0805.1726 [gr-qc]}.
\bibitem{10:DeFelice}A.~De~Felice and S.~Tsujikawa, ``{\em {$f(R)$
  Theories}},"\,\href{http://dx.doi.org/10.12942/lrr-2010-3}{Living Rev. Rel.
  {\bfseries 13}, 3 (2010)},
  \href{http://arxiv.org/abs/1002.4928}{arXiv:1002.4928 [gr-qc]}.
\bibitem{11:Nojiri}S.~Nojiri and S.~D. Odintsov, ``{\em {Unified cosmic history
  in modified gravity: From $F(R)$ theory to Lorentz non-invariant
  models}},"\,\href{http://dx.doi.org/10.1016/j.physrep.2011.04.001}{Phys.
  Rept. {\bfseries 505}, 59 (2011)},
  \href{http://arxiv.org/abs/1011.0544}{arXiv:1011.0544 [gr-qc]}.
\bibitem{07:Starobinsky}A.~A. Starobinsky, ``{\em {Disappearing cosmological
  constant in $f(R)$
  gravity}},"\,\href{http://dx.doi.org/10.1134/S0021364007150027}{JETP Lett.
  {\bfseries 86}, 157 (2007)},
  \href{http://arxiv.org/abs/0706.2041}{arXiv:0706.2041 [astro-ph]}.
\bibitem{07:Hu}W.~Hu and I.~Sawicki, ``{\em {Models of $f(R)$ cosmic
  acceleration that evade solar system
  tests}},"\,\href{http://dx.doi.org/10.1103/PhysRevD.76.064004}{Phys. Rev. D
  {\bfseries 76}, 064004 (2007)},
  \href{http://arxiv.org/abs/0705.1158}{arXiv:0705.1158 [astro-ph]}.
\bibitem{08:Cognola}G.~Cognola, E.~Elizalde, S.~Nojiri, S.~D. Odintsov,
  L.~Sebastiani, and S.~Zerbini, ``{\em {Class of viable modified $f(R)$
  gravities describing inflation and the onset of accelerated
  expansion}},"\,\href{http://dx.doi.org/10.1103/PhysRevD.77.046009}{Phys. Rev.
  D {\bfseries 77}, 046009 (2008)},
  \href{http://arxiv.org/abs/0712.4017}{arXiv:0712.4017 [hep-th]}.
\bibitem{07:Briscese}F.~Briscese, E.~Elizalde, S.~Nojiri, and S.~D. Odintsov,
  ``{\em {Phantom scalar dark energy as modified gravity: Understanding the
  origin of the Big Rip
  singularity}},"\,\href{http://dx.doi.org/10.1016/j.physletb.2007.01.013}{Phys.
  Lett. B {\bfseries 646}, 105 (2007)},
  \href{http://arxiv.org/abs/hep-th/0612220}{arXiv:hep-th/0612220}.
\bibitem{08:Nojiri}S.~Nojiri and S.~D. Odintsov, ``{\em {Future evolution and
  finite-time singularities in $F(R)$ gravity unifying inflation and cosmic
  acceleration}},"\,\href{http://dx.doi.org/10.1103/PhysRevD.78.046006}{Phys.
  Rev. D {\bfseries 78}, 046006 (2008)},
  \href{http://arxiv.org/abs/0804.3519}{arXiv:0804.3519 [hep-th]}.
\bibitem{08:Frolov}A.~V. Frolov, ``{\em {Singularity Problem with $f(R)$ Models
  for Dark
  Energy}},"\,\href{http://dx.doi.org/10.1103/PhysRevLett.101.061103}{Phys.
  Rev. Lett. {\bfseries 101}, 061103 (2008)},
  \href{http://arxiv.org/abs/0803.2500}{arXiv:0803.2500 [astro-ph]}.
\bibitem{08:Kobayashi}T.~Kobayashi and K.-i. Maeda, ``{\em {Relativistic stars
  in $f(R)$ gravity, and absence
  thereof}},"\,\href{http://dx.doi.org/10.1103/PhysRevD.78.064019}{Phys. Rev. D
  {\bfseries 78}, 064019 (2008)},
  \href{http://arxiv.org/abs/0807.2503}{arXiv:0807.2503 [astro-ph]}.
\bibitem{08:Dev}A.~Dev, D.~Jain, S.~Jhingan, S.~Nojiri, M.~Sami, and
  I.~Thongkool, ``{\em {Delicate $f(R)$ gravity models with a disappearing
  cosmological constant and observational constraints on the model
  parameters}},"\,\href{http://dx.doi.org/10.1103/PhysRevD.78.083515}{Phys.
  Rev. D {\bfseries 78}, 083515 (2008)},
  \href{http://arxiv.org/abs/0807.3445}{arXiv:0807.3445 [hep-th]}.
\bibitem{09:Kobayashi}T.~Kobayashi and K.-i. Maeda, ``{\em {Can higher
  curvature corrections cure the singularity problem in $f(R)$
  gravity?}},"\,\href{http://dx.doi.org/10.1103/PhysRevD.79.024009}{Phys. Rev.
  D {\bfseries 79}, 024009 (2009)},
  \href{http://arxiv.org/abs/0810.5664}{arXiv:0810.5664 [astro-ph]}.
\bibitem{10:Appleby}S.~A. Appleby, R.~A. Battye, and A.~A. Starobinsky, ``{\em
  {Curing singularities in cosmological evolution of $F(R)$
  gravity}},"\,\href{http://dx.doi.org/10.1088/1475-7516/2010/06/005}{J.
  Cosmol. Astropart. Phys. {\bfseries 1006}, 005 (2010)},
  \href{http://arxiv.org/abs/0909.1737}{arXiv:0909.1737 [astro-ph.CO]}.
\bibitem{60:Landau}L.~D. Landau and E.~M. Lifshitz, ``{\em {Mechanics}},"\,
  (Pergamon Press, New York, 1960).
\bibitem{16:Blas}D.~Blas, D.~L. Nacir, and S.~Sibiryakov, ``{\em {Ultralight
  Dark Matter Resonates with Binary
  Pulsars}},"\,\href{http://dx.doi.org/10.1103/PhysRevLett.118.261102}{Phys.
  Rev. Lett. {\bfseries 118}, 261102 (2017)},
  \href{http://arxiv.org/abs/1612.06789}{arXiv:1612.06789 [hep-ph]}.
\end{thebibliography}

\end{document}